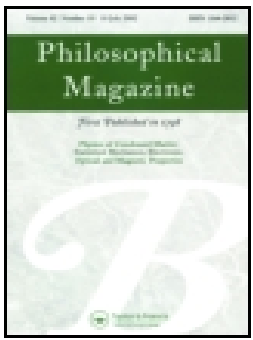

# Infrared spectroscopy of hydrogenated and chlorinated amorphous silicon

S. Kalem , R. Mostefaoui , J. Bourneix & J. Chevallier

# Infrared spectroscopy of hydrogenated and chlorinated amorphous silicon

By S. Kalem, R. Mostefaoui, J. Bourneix
and J. Chevallier

Laboratoire de Physique des Solides, C.N.R.S.,
1 place A. Briand, 92195 Meudon Principal, France

## Abstract

The evolution is presented of infrared transmission spectra of hydrogenated and chlorinated amorphous silicon films versus the chlorine concentration and the annealing temperature. As the chlorine content increases, an increase of the $2100\,\text{cm}^{-1}$ band intensity is observed. We associate this effect with the existence of $SiH_xCl_y$ mixed configurations. The SiCl stretching band, located at $545\,\text{cm}^{-1}$ in unannealed samples, shifts continuously up to $590\,\text{cm}^{-1}$ as the annealing temperature increases. This displacement is attributed to a predominance of $SiCl_2$ species around $500°C$ and their transformation to $SiCl_3$ species and $SiCl_4$ molecules above $700°C$.

## § 1. Introduction

Fundamental studies of the properties of hydrogenated amorphous silicon (a-Si:H) have been widely developed in the recent past, partly because of its applications in the field of photovoltaic conversion of solar energy. The total or partial replacement of hydrogen by halogen atoms has been proposed more recently. Fluorinated and hydrogenated amorphous silicon already exhibits a promising performance as illustrated by the achievement of solar cells with 9% efficiency (Hamakawa 1982). Chlorinated and hydrogenated amorphous silicon (a-Si:H,Cl) has also received some attention and the first fundamental studies on the material have been performed (Kruehler, Plaettner, Moeller, Rauscher and Stetter 1980, Fortunato, Evangelisti, Bruno, Capezutto, Cramarossa, Augelli and Murri 1981, Chevallier, Kalem, Al Dallal and Bourneix 1982, Danish, Georgiev and Jahn 1984). It has been shown, for example, that chlorine could be a dangling bond compensator, as are hydrogen and fluorine, in an amorphous silicon matrix. The transport and recombination properties of a-Si:H depend on the nature of the $SiH_n$ groupings present in the amorphous silicon matrix. For this reason, a detailed study of the nature of the groupings ($SiH_n$, $SiCl_n$, SiHCl) present in a-Si:H,Cl is important for a good understanding of the material properties. In this paper, we present the vibrational properties of a-Si:H,Cl films containing various amounts of chlorine. We shall discuss the origin of the absorption bands with the help of annealing experiments.

## § 2. Preparation

Hydrogenated and chlorinated amorphous silicon films were prepared by decomposition of a $SiCl_4 + H_2$ mixture in a r.f. glow discharge. In these experiments, we used a total pressure of 0·5 Torr. The r.f. power is 5 W and the substrate temperature is 250°C.

Fig. 1

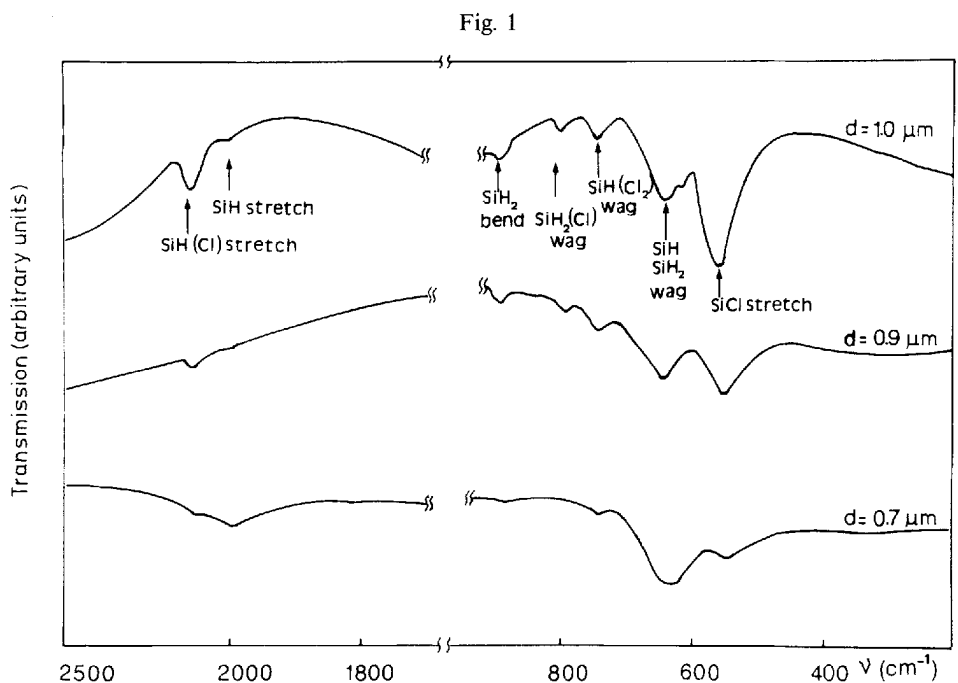

Infrared transmission spectra of a-Si: H,Cl layers prepared using liquid SiCl$_4$ at temperatures of +6°C (8·9% H and 4·2% Cl) (upper graph), +11°C (4·2% H and 7·4% Cl) (middle graph) and +30°C (4·0% H and 11·2% Cl) (lower graph).

All these parameters were the same for all the films. The only parameter deliberately changed was the temperature of the liquid SiCl$_4$, which was varied in the range −15°C–30°C. The subsequent variation of SiCl$_4$ partial pressure modifies the SiCl$_4$/H$_2$ partial pressure ratio and thus the amount of chlorine in the films. Using this parameter, we have been able to vary the chlorine concentration in the range 2–11% as determined by electron microprobe analysis (Mostefaoui, Chevallier, Mechenin and Auzel 1984).

### §3. INFRARED SPECTROSCOPY

Infrared transmission spectra of a-Si: H,Cl films have been performed in the range 4000–200 cm$^{-1}$. The layers were deposited on high-resistivity crystalline silicon substrates.

Figure 1 presents the spectra of several a-Si: H,Cl films prepared using liquid SiCl$_4$ at different temperatures. All these spectra show the existence of the same absorption bands with different relative intensities.

#### 3.1. SiH induced bands

The origin of the bands at 1990, 2110, 890, 840 and 640 cm$^{-1}$ is quite clear. Referring to the vibrational properties of a-Si: H, these bands are attributed to the stretching modes of Si–H (1990 cm$^{-1}$) and Si–H$_2$ (2110 cm$^{-1}$), the bending modes of SiH$_2$ groups (890 cm$^{-1}$) and (SiH$_2$)$_n$ chains (890 and 840 cm$^{-1}$), and the wagging–rocking mode of SiH and SiH$_2$ species (640 cm$^{-1}$) (Brodsky, Cardona and Cuomo 1977). However, we notice an increase in the 2110 cm$^{-1}$ band intensity as the band at 545 cm$^{-1}$, which due to the stretching vibration mode of SiCl species (Kalem,

Chevallier, Al Dallal and Bourneix 1981), increases in intensity. Moreover, we do not observe any significant variation of the 840 and 890 cm$^{-1}$ band intensities, which means that the amount of SiH$_2$ species remains unchanged as the chlorine concentration increases. The increase of the 2110 cm$^{-1}$ band intensity can be readily explained by assuming a significant contribution of SiH(Cl) mixed groupings to this band as we have proposed previously (Kalem *et al.* 1981). The reason for this can be provided in terms of an induction effect of chlorine atoms present in such groups. The stretching-mode frequency of the SiH bonds is known to increase as the electronegativity of the next-nearest-neighbour atom increases (Lucovsky 1979). So the presence of a chlorine atom in SiH(Cl) groups implies an increase of the SiH stretching-mode frequency.

In a previous paper, we assigned the bands at 740 and 795 cm$^{-1}$ to SiH vibrational modes from deuteration experiments (Chevallier, Kalem, Bourneix and Vandevyver 1983). Moreover, annealing experiments show that the 740 cm$^{-1}$ band is thermally more stable than the 795 cm$^{-1}$ band (fig. 2). This result indicates that these two bands

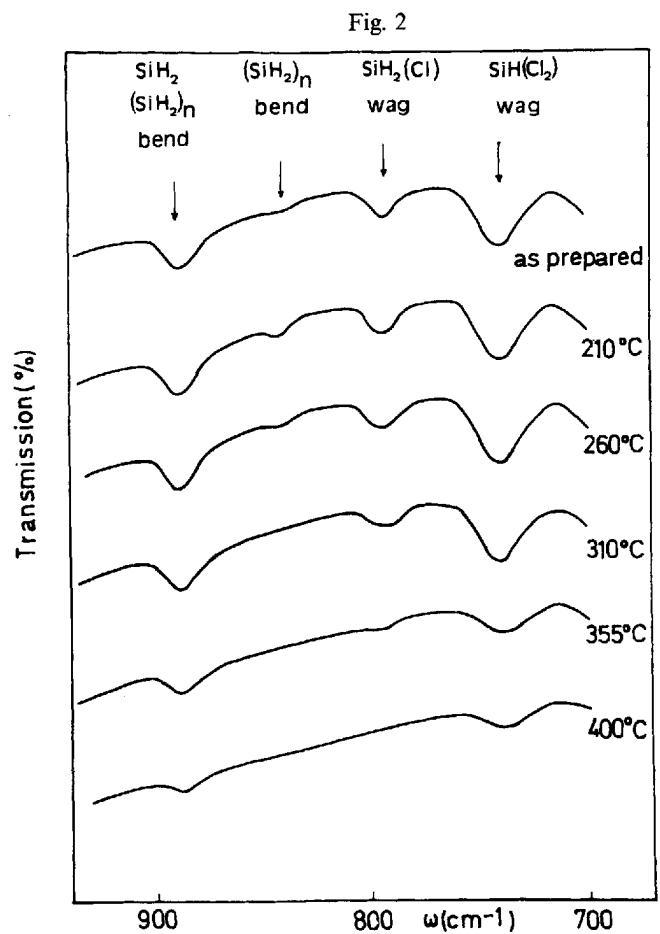

Evolution of the infrared transmission spectrum of an a-Si:H,Cl sample with annealing temperature in the range 700–900 cm$^{-1}$.

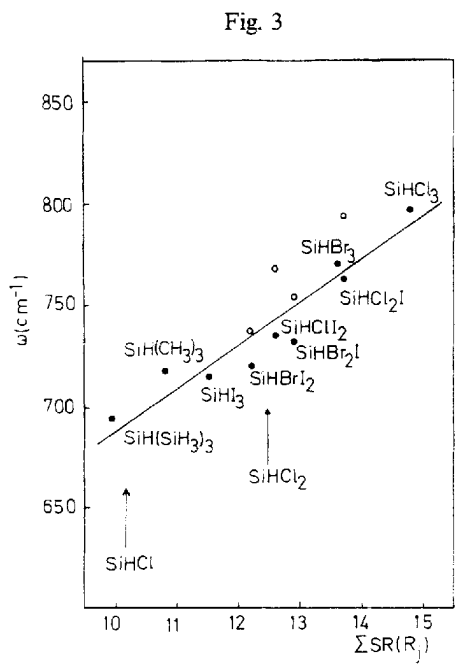

Fig. 3

Variation of the Si–H wagging-mode frequency with the electronegativity sum of X and X' atoms in $HSi(X)_3$ and $HSiX(X')_2$ molecules.

originate from two different species. We have first studied the possibility of assigning these two infrared bands to molecules of $SiH_nCl_{4-n}$ types ($n = 1, 2, 3$) embedded in the matrix. According to the data in the literature, no molecule of this type has an absorption band at 740 cm$^{-1}$, but the wagging mode of Si–H in $SiHCl_3$ is at 800 cm$^{-1}$, and so the 795 cm$^{-1}$ band might be attributed to such molecules. However, the Si–H wagging band in $SiHCl_3$ at 800 cm$^{-1}$ has an intensity as high as the Si–H stretching mode at 2200 cm$^{-1}$, which was not observed in the spectra. Therefore, we exclude $SiHCl_3$ molecules from our as-prepared a-Si:H,Cl layers.

In fig. 3 we report the evolution of the Si–H wagging-mode frequencies versus the electronegativity sum of radicals adjacent to the Si–H bonds (Kolessova, Kukharskaja and Andreev 1953, Hengge and Höfler 1971, Brodsky *et al.* 1977). As already known for the Si–H stretching mode, we observe an increase in the Si–H wagging-mode frequency as the electronegativity of the adjacent radicals increases. Figure 4 shows that the same dependence occurs for the twisting and the rocking modes of $SiH_2$ groups in $SiH_2XX'$ molecules (Ebsworth, Onyszchuk and Sheppard 1958). The bending mode does not seem as sensitive as the other modes to the characteristics of the adjacent radicals. In the case of the wagging mode of $SiH_2$ species, the influence of the electronegativity is less clear. However, we are able to conclude from these observations that the 740 and 795 cm$^{-1}$ bands are due to Si–H-related deformation modes of two different species containing hydrogen and chlorine, displaced from 640 cm$^{-1}$ towards high frequencies by induction effects due to the presence of chlorine.

We note from fig. 3 that the electronegativity increase due to the replacement of one silicon atom by one chlorine atom would induce an increase of 50 cm$^{-1}$ in the wagging-mode frequency of SiH radicals. For this reason, we suggest that the 740 cm$^{-1}$ band

Fig. 4

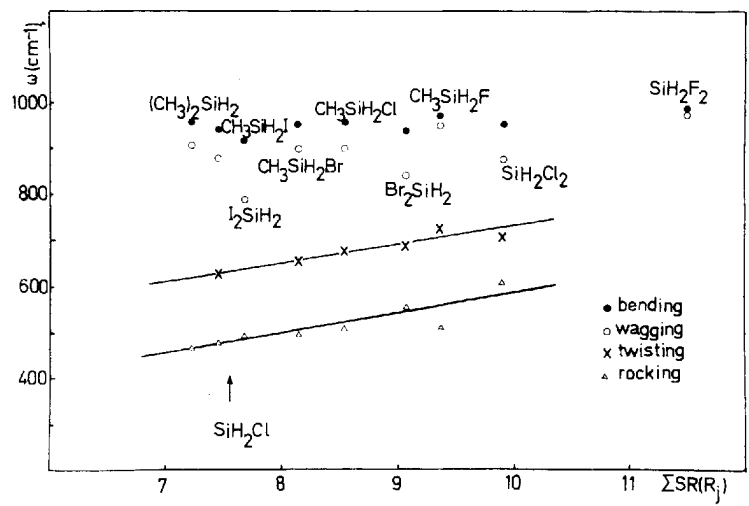

Variation of the bending-, wagging-, twisting- and rocking-mode frequencies of $SiH_2$ with the electronegativity sum of X and X' radicals in $SiH_2X_2$ and $SiH_2XX'$ molecules.

originates from the wagging mode of Si–H in $SiHCl_2$ configurations. The discussion of the 795 cm$^{-1}$ band is less easy. However, we see from fig. 4 that this frequency falls in the range of the wagging-mode frequencies of $SiH_2$ radicals. So we propose to assign the 795 cm$^{-1}$ band to the wagging mode of $SiH_2$ in $SiH_2Cl$ configurations. The assignment of the 740 and 795 cm$^{-1}$ bands is quite consistent with our previous conclusion as to the existence of mixed configurations in the matrix.

In the case of $SiHCl_2$ species, we would expect the existence of a second Si–H wagging mode. The reason can be derived simply from symmetry considerations. The $SiHX_3$ types of molecules or species have $C_{3v}$ symmetry and the Si–H wagging mode is doubly degenerate (E-type symmetry). Replacing one of the X atoms by an X' gives the species $C_s$ symmetry and, as a consequence of this symmetry lowering, the Si–H wagging mode is split into two modes of symmetry, $A_1$ and $A_2$. However, the second component of this wagging mode has not been detected, either because of the small splitting value or because of the relative weakness of the oscillator strength of the optical transition associated to this second mode.

### 3.2. SiCl induced bands

In addition to the SiH bands, fig. 1 shows the existence of two other bands located at 545 and 615 cm$^{-1}$. In a previous paper, we demonstrated that these two bands were unaffected by deuteration and consequently have to be attributed to SiCl vibrational modes (Chevallier et al. 1983). In order to clarify the origin of these two bands, we performed on our samples a series of isochronal thermal annealings in a $H_2$ atmosphere. An annealing time of 30 min was chosen, and fig. 5 shows the typical evolution of the absorption spectrum with the annealing temperature $T_a$. Figure 6 shows the integrated intensity of each band versus $T_a$.

All the hydrogen band intensities start decreasing above 300°C and disappear at 600°C. We notice that the 740 and 795 cm$^{-1}$ bands behave similarly to the other

Fig. 5

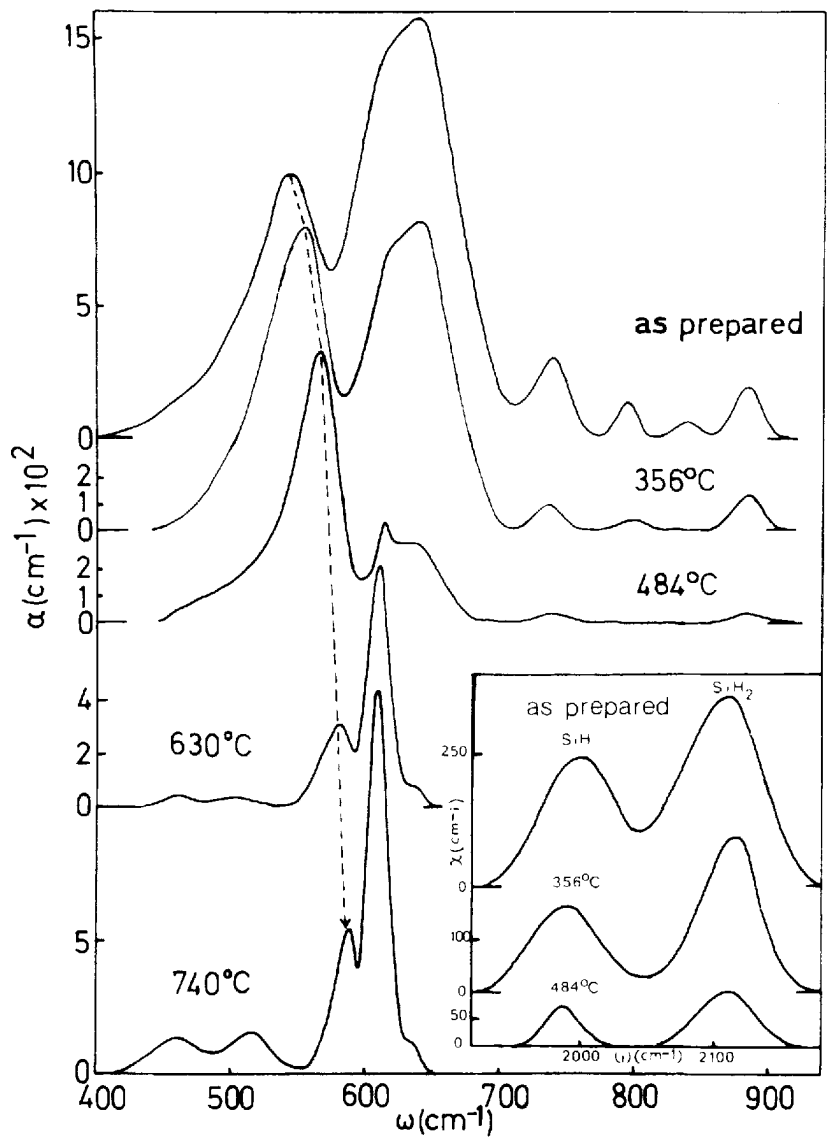

Evolution of the infrared transmission spectrum of an a-Si:H,Cl sample with the annealing temperature.

hydrogen bands. The intensity of the $615\,\text{cm}^{-1}$ band starts decreasing at $300°\text{C}$ and completely vanishes above $400°\text{C}$. The $545\,\text{cm}^{-1}$ band intensity decreases above $350°\text{C}$.

A remarkable feature of the $545\,\text{cm}^{-1}$ band is its continuous shift from 545 to $590\,\text{cm}^{-1}$ as the annealing temperature increases from $300°\text{C}$ to $700°\text{C}$, as shown in fig. 5 and 7. In fig. 5 we observe a decrease of the $615\,\text{cm}^{-1}$ band intensity followed by an increase in this band. We start the discussion on the origin of the 545 and $615\,\text{cm}^{-1}$ bands by looking at the table. This table provides the SiCl stretching mode frequencies

Fig. 6

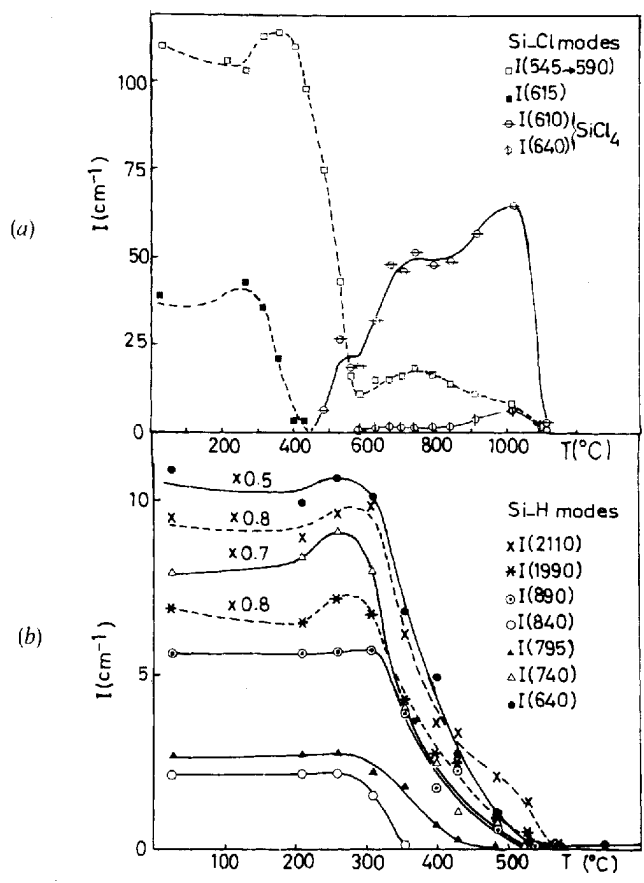

Integrated absorption of the (a) SiCl and (b) SiH related bands versus the annealing temperature.

deduced from the electronegativity approach which is based on the linear correlation existing between the SiCl stretching-mode frequency and the electronegativity sum of adjacent radicals (Kalem et al. 1981). Apart from the symmetric mode frequency of the SiCl$_3$ group at 455 cm$^{-1}$, all the SiCl stretching-mode frequencies shown in this table are located between 495 and 600 cm$^{-1}$. We note from the values reported in the table that the 545 cm$^{-1}$ band can originate from several kinds of configurations. Among the different possibilities, we first mention that this band could correspond to the symmetric stretching mode of Si–Cl$_2$ in (SiCl$_2$)$_n$ polymers. The corresponding asymmetric mode is expected at 598 cm$^{-1}$, which is the closest value to 615 cm$^{-1}$. So, from these induction considerations, the 615 cm$^{-1}$ band and a part of the 545 cm$^{-1}$ band would be attributed to (SiCl$_2$)$_n$ polymer groups which disappear under annealing above 400°C.

According to the data in the literature, the asymmetric branch of the Si–Cl$_n$ vibrations in SiCl$_n$X$_{4-n}$ (X = H, Br, Cl, F..., n = 2, 3) and in SiCl$_n$(CH$_3$)$_{4-n}$ (n = 2, 3) has always a higher intensity than the symmetric band. For example, in the SiH$_2$Cl$_2$ and (CH$_3$)$_2$SiCl$_2$ molecules, the oscillator strength (Γ) of the Si–Cl$_2$ asymmetric stretching

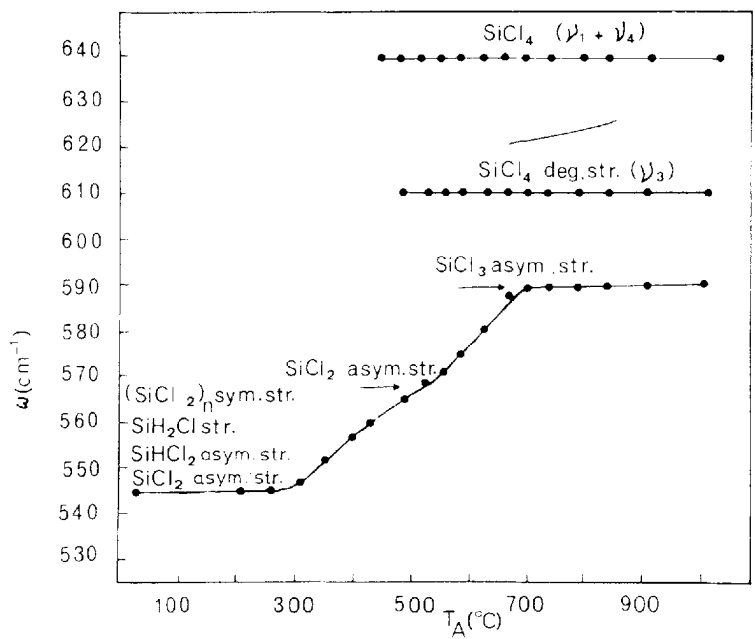

Fig. 7

Displacement of the '545 cm$^{-1}$ band' as a function of the annealing temperature.

mode is approximately 11 cm$^2$ mmol$^{-1}$ and that of the symmetric mode is about 3 cm$^2$ mmol$^{-1}$. Here we suppose that the same ratio of symmetric and asymmetric stretching frequencies oscillator strength is valid for SiCl$_2$ in our a-Si : H,Cl films. In the spectrum of the sample shown in fig. 5, the integrated absorption strength of the band at 615 cm$^{-1}$ is typically 25 cm$^{-1}$. In these conditions, we deduce a value of 7 cm$^{-1}$ for the integrated absorption strength of the Si–Cl$_2$ symmetric stretching mode at 545 cm$^{-1}$. In the absorption spectra, the intensity of the 545 cm$^{-1}$ band is of the order of 75 cm$^{-1}$. This result confirms that the band at 545 cm$^{-1}$ is the superposition of several absorption bands due to various kinds of configurations.

We note from the table that the Si–Cl stretching-mode frequency in SiH$_2$Cl species is close to the experimental frequency at 545 cm$^{-1}$. Moreover, the presence of SiH$_2$Cl species has already been proved from the observation of a band at 795 cm$^{-1}$ attributed to the wagging mode of SiH$_2$ radicals in SiH$_2$Cl groupings. Similarly, the observation of the 740 cm$^{-1}$ band has been correlated to the Si–H wagging mode of SiHCl$_2$ groupings. For these reasons, we expect that the Si–Cl stretching mode in SiH$_2$Cl species and of the Si–Cl$_2$ asymmetric stretching mode in SiHCl$_2$ species also contribute to the 545 cm$^{-1}$ band.

The interesting feature already mentioned concerns the shift of the chlorine-related bands at 545 cm$^{-1}$ towards high frequencies with increasing annealing temperature (fig. 7). The frequency variation shows a first kink at 570 cm$^{-1}$ around 500°C and reaches a saturation value of 590 cm$^{-1}$ above 700°C. Above 600°C, the shift is accompanied by an increase of the intensity of this band and the appearance of two new bands located at 610 and 640 cm$^{-1}$ (figs. 5 and 6). All these bands vanish at 1100°C. We interpret the behaviour of the infrared spectra above 500°C as the result of the SiCl stretching-mode frequencies in various species as deduced from the electronegativity measurements.

| Species | $v_s$(SiCl) (cm$^{-1}$) |
|---|---|
| SiCl | 523 |
| SiClH | 532 |
| SiClH$_2$ | 541 |
| SiCl$_2$H | 510, 577 |
| SiCl$_2$ | 495, 567 |
| SiCl$_3$ | 455, 588 |
| (SiCl$_2$)$_n$ | 544, 598 |

formation of SiCl$_3$ radicals and SiCl$_4$ molecules embedded in the amorphous silicon matrix.

The table shows that the asymmetric stretching mode of SiCl$_3$ radicals is expected at 588 cm$^{-1}$, in good agreement with the experimental value of 590 cm$^{-1}$. The corresponding symmetric stretching mode would be at 455 cm$^{-1}$ and falls in the phonon sideband of the matrix. In fig. 5 we observe an increase of the absorption intensity in this region which is indicative of the existence of this symmetric mode. Literature data indicate that the 610 and 640 cm$^{-1}$ bands can be interpreted as the frequencies of the $v_3$ degenerate stretching band and the $v_1+v_4$ band of SiCl$_4$ molecules. The above results show that a formation of SiCl$_3$ radicals and SiCl$_4$ molecules occurs in the matrix at elevated temperatures. Moreover, electron microprobe analysis shows that the chlorine concentration in the layers remains almost constant up to annealing temperatures of 1000°C. At 1100°C, no chlorine is detected. This result means that, contrary to hydrogen which exodiffuses, chlorine lies inside the matrix, which probably, being large, migrates less easily than hydrogen. However, chlorine atoms migrate enough to give rise to new SiCl$_3$ radicals and SiCl$_4$ molecules.

The shift of the 545 cm$^{-1}$ band and the decreases of the 615 cm$^{-1}$ band intensity and the various hydrogen related band intensities simultaneously start at about 300°C. We suggest that above this temperature, hydrogen is progressively released from the SiH$_n$ and SiH$_x$Cl$_y$ configurations and partly replaced by chlorine. Moreover, the (SiCl$_2$)$_n$ chains are broken. These two mechanisms would favour the formation of SiCl$_2$ and SiCl$_3$ radicals and also SiCl$_4$ molecules. The shift of the 545 cm$^{-1}$ band between 300 and 480°C is attributed to an increasing predominance of isolated SiCl$_2$ species in the '545 cm$^{-1}$ band'. At 500°C, this predominance is seen as a slight saturation value at 570 cm$^{-1}$ of the main chlorine band, corresponding to the asymmetric stretching mode of isolated SiCl$_2$ species, in good agreement with the induction effect predictions. The corresponding symmetric stretching mode of these SiCl$_4$ species is expected at 495 cm$^{-1}$. Again, this is in good agreement with the experimental observation of a band at 515 cm$^{-1}$ close to the phonon side band of the matrix.

### §4. Hydrogen and chlorine concentration

The total hydrogen content of unannealed samples has been measured by the nuclear reaction method. We have plotted the integrated strength of the two SiH infrared stretching bands at 1990 and 2110 cm$^{-1}$ and that of the SiH wagging band at 640 cm$^{-1}$ versus the total hydrogen concentration $C$(H) measured on the same samples. The relationships between $I_s$, $I_w$ and $C$(H) can be approximated by a straight

518     S. Kalem *et al.*

line. The proportionality constants $A_s$ and $A_w$, such that $N_H = AI$ where $N_H$ is the number of hydrogen atoms per cubic centimetre, have the following numerical values:

$$A_s = 1\cdot 4 \times 10^{20} \text{ atoms cm}^{-2},$$

$$A_w = 1\cdot 8 \times 10^{19} \text{ atoms cm}^{-2}.$$

These values agree well with the corresponding values for unchlorinated a-Si:H samples obtained earlier (Shanks, Fang, Cardona, Demond and Kalbitzer 1980). Moreover, Fang *et al.* have shown that the proportionality constant $A_w$ is the same in a-Si:H,F as in a-Si:H (Fang, Ley, Shanks, Gruntz and Cardona 1980). As a consequence, the oscillator strength of the wagging and stretching modes of Si–H bonds is not significantly influenced by the presence of other halogen elements such as Cl and F.

The chlorine concentrations of the unannealed films were determined by electron microprobe analysis and nuclear activation. In fig. 8 we see that the integrated absorption strength of the SiCl stretching band at 545 cm$^{-1}$ is a linear function of the

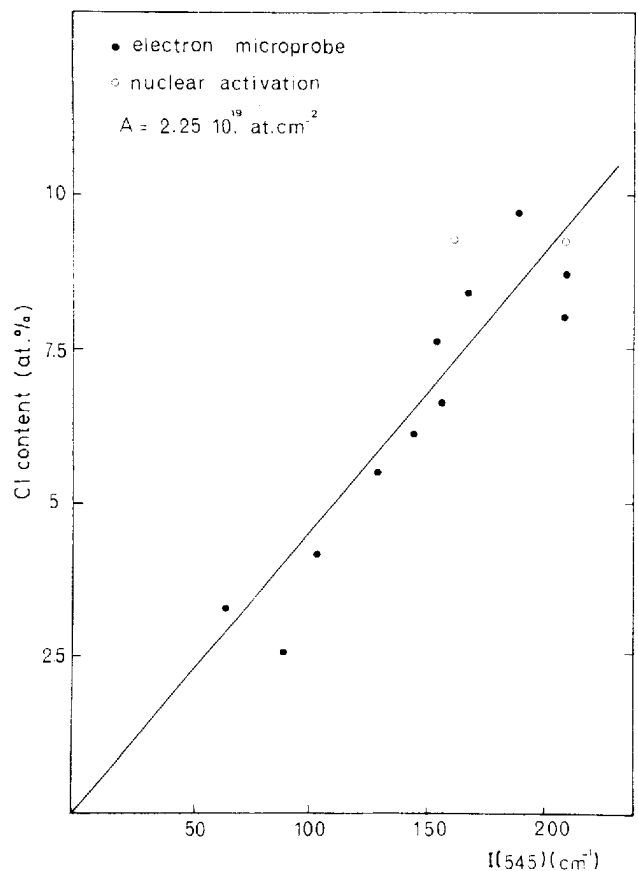

Fig. 8

Variation of the chlorine content measured by electron microprobe and nuclear activation analysis as a function of the integrated infrared absorption band at 545 cm$^{-1}$.

measured chlorine concentration for a series of samples containing from 1·6% to $\simeq 10$ at.% chlorine. The numerical value of the proportionality constant is

$$A_s(\text{Si–Cl}) = 2 \cdot 25 \times 10^{19} \text{ atoms cm}^{-2}.$$

Wu Zhi Quiang *et al.* found that the extrapolation of the $I = f(\%\text{Cl})$ line gives a non-zero intercept on the chlorine concentration axis (Wu Zhi Quiang, Xu Cun-yi, Zhang Wei-ping, Zhen Zhao-bo and Fang Rong-chuan 1983). They deduced that isolated $SiCl_4$ molecules would be trapped in the matrix of their unannealed samples. From fig. 8 we have no clear evidence of these phenomena in our unannealed samples: $SiCl_4$ molecules are only present under annealing above 500°C.

### § 5. Conclusion

Using infrared transmission spectroscopy, we have been able to deduce several conclusions on the origin of the various SiH and SiCl-related absorption bands in our a-Si:H,Cl films. The existence of several SiH related bands (740, 795, 2110 cm$^{-1}$) induced by an electronegativity effect is evidence of the presence of a significant amount of $SiH_xCl_y$ configurations in our material. The shift of the SiCl stretching band with annealing temperature showed the predominance of $SiCl_2$ species in samples annealed around 500°C and their transformation in $SiCl_3$ species and $SiCl_4$ molecules for annealing temperatures above 700°C. Contrary to hydrogen in a-Si:H which exodiffuses above 300°C, chlorine remains in a-Si:H,Cl up to 1000°C.

In a-Si:H the correlation found by several authors between the infrared spectra and other properties of the material (spin density, electronic density of states) has been one of the bases for an improvement in the material quality. For this reason it would be interesting to try to correlate the above basic results with other physical properties, especially the electronic density of states in the band gap of our material.

### Acknowledgments

We thank Mrs M. Rommeluere for the electron microprobe analysis and M. Toulemonde, of CRN Strasbourg, for the nuclear activation analysis.